\documentstyle[aaspp4]{article}	
\makeatletter
\let\@internalcite\cite
\def\cite{\def\@citeseppen{-1000}%
    \def\@cite##1##2{(##1\if@tempswa , ##2\fi)}%
    \def\citeauthoryear##1##2##3{##1 ##3}\@internalcite}

\def\citeNP{\def\@citeseppen{-1000}%
    \def\@cite##1##2{##1\if@tempswa , ##2\fi}%
    \def\citeauthoryear##1##2##3{##1 ##3}\@internalcite}
\def\citeN{\def\@citeseppen{-1000}%
    \def\@cite##1##2{##1\if@tempswa , ##2)\else{)}\fi}%
    \def\citeauthoryear##1##2##3{##1 (##3}\@citedata}
\def\citeA{\def\@citeseppen{-1000}%
    \def\@cite##1##2{(##1\if@tempswa , ##2\fi)}%
    \def\citeauthoryear##1##2##3{##1}\@internalcite}
\def\citeANP{\def\@citeseppen{-1000}%
    \def\@cite##1##2{##1\if@tempswa , ##2\fi}%
    \def\citeauthoryear##1##2##3{##1}\@internalcite}
\def\shortcite{\def\@citeseppen{-1000}%
    \def\@cite##1##2{(##1\if@tempswa , ##2\fi)}%
    \def\citeauthoryear##1##2##3{##2 ##3}\@internalcite}
\def\shortciteNP{\def\@citeseppen{-1000}%
    \def\@cite##1##2{##1\if@tempswa , ##2\fi}%
    \def\citeauthoryear##1##2##3{##2 ##3}\@internalcite}
\def\shortciteN{\def\@citeseppen{-1000}%
    \def\@cite##1##2{##1\if@tempswa , ##2)\else{)}\fi}%
    \def\citeauthoryear##1##2##3{##2 (##3}\@citedata}
\def\shortciteA{\def\@citeseppen{-1000}%
    \def\@cite##1##2{(##1\if@tempswa , ##2\fi)}%
    \def\citeauthoryear##1##2##3{##2}\@internalcite}
\def\shortciteANP{\def\@citeseppen{-1000}%
    \def\@cite##1##2{##1\if@tempswa , ##2\fi}%
    \def\citeauthoryear##1##2##3{##2}\@internalcite}
\def\citeyear{\def\@citeseppen{-1000}%
    \def\@cite##1##2{(##1\if@tempswa , ##2\fi)}%
    \def\citeauthoryear##1##2##3{##3}\@citedata}
\def\citeyearNP{\def\@citeseppen{-1000}%
    \def\@cite##1##2{##1\if@tempswa , ##2\fi}%
    \def\citeauthoryear##1##2##3{##3}\@citedata}

\def\@citedata{%
	\@ifnextchar [{\@tempswatrue\@citedatax}%
				  {\@tempswafalse\@citedatax[]}%
}

\def\@citedatax[#1]#2{%
\if@filesw\immediate\write\@auxout{\string\citation{#2}}\fi%
  \def\@citea{}\@cite{\@for\@citeb:=#2\do%
    {\@citea\def\@citea{, }\@ifundefined
       {b@\@citeb}{{\bf ?}%
       \@warning{Citation `\@citeb' on page \thepage \space undefined}}%
{\csname b@\@citeb\endcsname}}}{#1}}%

\def\@citex[#1]#2{%
\if@filesw\immediate\write\@auxout{\string\citation{#2}}\fi%
  \def\@citea{}\@cite{\@for\@citeb:=#2\do%
    {\@citea\def\@citea{; }\@ifundefined
       {b@\@citeb}{{\bf ?}%
       \@warning{Citation `\@citeb' on page \thepage \space undefined}}%
{\csname b@\@citeb\endcsname}}}{#1}}%

\def\@biblabel#1{}

\newlength{\bibhang}
\makeatother

\def\kms{km~s$^{-1}$}

\def\htwo{H$_2$}

\def\orf{ORFEUS}
\def\hxi{\ion{H}{1}}
\def\hexi{\ion{He}{1}}
\def\oxi{\ion{O}{1}}
\def\nxi{\ion{N}{1}}

\def\nxiii{\ion{N}{3}}
\def\cxii{\ion{C}{2}}
\def\cxiii{\ion{C}{3}}
\def\fexii{\ion{Fe}{2}}
\def\fexiii{\ion{Fe}{3}}
\def\oxvi{\ion{O}{6}}
\def\nxv{\ion{N}{5}}
\def\cxiv{\ion{C}{4}}
\def\arxi{\ion{Ar}{1}}
\def\pxii{\ion{P}{2}}
\def\sxii{\ion{S}{2}}
\def\sxiii{\ion{S}{3}}
\def\sixiii{\ion{Si}{3}}
\def\clxi{\ion{Cl}{1}}

\begin{document}
\title{ORFEUS-II FAR-ULTRAVIOLET OBSERVATIONS OF 3C273:\\
1. INTERSTELLAR AND INTERGALACTIC ABSORPTION LINES\footnotemark}
\footnotetext{Based on the development and utilization of
\orf\ (Orbiting and Retrievable Far and Extreme Ultraviolet
Spectrometers), a collaboration of the Institute for Astronomy
and Astrophysics of the
University of T\"{u}bingen, the Space Astrophysics Group of the
University of California at Berkeley, and the Landessternwarte
Heidelberg.}


\author{Mark~Hurwitz, Immo~Appenzeller$^2$, Juergen~Barnstedt$^3$, Stuart~Bowyer,
W.~Van~Dyke~Dixon, Michael~Grewing$^3$, Norbert~Kappelmann$^3$, Gerhard~Kraemer$^3$,
Joachim~Krautter$^2$, and Holger~Mandel$^2$}

\affil{Space Sciences Laboratory\\
University of California, Berkeley, California 94720-7450}

\altaffiltext{2}{Landessternwarte, University of Heidelberg}
\altaffiltext{3}{Institute for Astronomy and Astrophysics, University of T\"{u}bingen}

\authoremail{markh@ssl.berkeley.edu}


\begin{abstract}

We present the first intermediate-resolution ($\lambda$ / 3000) 
spectrum of the bright quasi-stellar object 3C273
at wavelengths between 900 and 1200 \AA.
Observations were performed with the Berkeley
spectrograph aboard the \orf-II mission \cite{Hetal98}.
We detect Lyman $\beta$ counterparts to intergalactic Lyman $\alpha$
features identified by \citeN{MWSG91} at $cz$ = 19900, 1600, and 1000 \kms;
counterparts to other putative Lyman $\alpha$ clouds 
along the sight line are below our detection limit.
The strengths of the two very low redshift Lyman $\beta$ features,
which are believed to arise in Virgo intracluster gas, 
exceed preflight expectations \cite{WRWMH95},
suggesting that the previous determination of the
cloud parameters may underestimate the true column densities.
A curve-of-growth analysis sets a minimum \hxi\ column density 
of $4 \times 10^{14}$ cm$^{-2}$ for the 1600 \kms\ cloud.
We find marginally significant evidence for
Galactic \htwo\ along the sight line, with a
total column density of about $10^{15}$ cm$^{-2}$.
We detect the stronger interstellar \oxvi\ doublet 
member unambiguously; the weaker member is blended with other features.
If the Doppler $b$ value for \oxvi\ is comparable
to that determined for \nxv\ \cite{SST97} then
the \oxvi\ column density is $7 \pm 2 \times 10^{14}$ cm$^{-2}$,
significantly above the only previous estimate \cite{D93}.
The \oxvi / \nxv\ ratio is about 10, consistent with
the low end of the range observed in the 
disk (compilation of \citeNP{HB96}).
Additional interstellar species detected for the first time
toward 3C273 (at modest statistical significance)
include \pxii, \fexiii, \arxi, and \sxiii.

\end{abstract}
 
\keywords{ISM: atoms, galaxies: clusters: individual: Virgo, quasars: individual: 3C273}

\section{OBSERVATIONS}

We present the first intermediate-resolution
far-ultraviolet spectrum of the bright quasi-stellar
object 3C273 in the 900 -- 1210 \AA\ band,
emphasizing absorption features arising along the line of sight.
\citeN{Appetal98} discuss the intrinsic spectrum of 3C273.
These data were collected with the Berkeley
spectrograph \cite{Hetal98} during the ORFEUS-SPAS II mission.
The \orf\ project and the {\it ASTRO-SPAS\/} platform are 
described in \citeN{Getal91}.

3C273 was observed five times, for a total of 10,797 s.  
All observations took place in late 1996, between 
330/21:41 and 338/16:33 (Day of Year/HH:MM, GMT).
The target coordinates were $\alpha$ = 12 29 06.7, $\delta$ = +02 03 09 (J2000).
Absolute positioning of the 26$\arcsec$ diameter
ORFEUS entrance aperture is accurate to $\pm$ 5$\arcsec$.
Extraction of the spectrum and subtraction of airglow follow
the discussion in \citeN{Hetal98}.
Statistical and systematic uncertainties associated with 
background and airglow subtraction are properly tracked.
As a final step, we bin the data on 0.1 \AA\ centers.

In Figure 1 we show the complete spectrum (solid line),
the 1 $\sigma$ uncertainty associated with shot noise
and detector flat-field effects (dotted line), 
and a continuum established
by an automated fitting routine (dashed line).  The automated
fitting routine identifies absorption features deeper than
a statistically determined threshold, replaces those data with 
a linear interpolation from nearby wavelengths, smoothes heavily, 
then iterates twice more at increasing sensitivity to absorption lines.
The resulting continuum has the advantage of being determined 
objectively, and evidently tracks most of the true spectrum well at 
wavelengths greater than 1000 \AA.  
Like many such routines, however, this one
yields unreliable results when strong features are closely 
spaced on a curved continuum (as in the region of absorption
lines 3 and 4) and will tend to underestimate the true
value when the spectrum contains a cluster of
weak absorption features below the statistical threshold
(note the weak dip in the fitted continuum around 1050 \AA).

In Table 1 we list the interstellar absorption features and 
blends that are sufficiently strong and isolated for direct equivalent 
width measurement using the continuum shown in Figure 1
(unless otherwise noted).
Statistical noise in the spectrum (1$\sigma$) corresponds 
to an unresolved absorption feature with equivalent 
width of about 0.040 \AA;
continuum placement uncertainty corresponds to about half that value.  
The value in the fourth column includes the statistical uncertainty only.
For species whose column densities have
not previously been reported, we list the minimum column
(if optically thin) based on the best fit equivalent width and the
oscillator strenths of \citeN{M91}.

\begin{deluxetable}{llllll}
\tablewidth{0pt}
\tablecaption{Interstellar Lines and Equivalent Widths}

\tablehead{
\colhead{\#}            &
\colhead{W$_{obs}$ (\AA)}           &
\colhead{W$_{eq}$ (\AA)}           &
\colhead{$\sigma$ (\AA)}      &
\colhead{ID}        &
\colhead{Log ($N_{min}$) (cm$^{-2}$)}
}
\startdata

1 & 1206.44 & 0.573 & 0.056 & Si III 1206.50 & 13.4 \nl
2 & 1199.75 & 0.896 & 0.114 & N I blend & \nl
3\tablenotemark{a} & 1193.56 & 0.510 & 0.050 & Si II 1193.29 & \nl
4\tablenotemark{a} & 1190.58 & 0.532 & 0.051 & Si II 1190.42 & \nl
5 & 1152.91 & 0.105 & 0.038 & P II 1152.82 & 13.6 \nl
6 & 1144.92 & 0.340 & 0.054 & Fe II 1144.94 & \nl
7 & 1143.44 & 0.137 & 0.040 & Fe II 1143.22 & \nl
8 & 1134.63 & 1.135 & 0.084 & N I blend & 15.1 \nl
9 & 1122.54 & 0.157 & 0.038 & Fe III 1122.53 & 14.3 \nl
10 & 1121.99 & 0.149 & 0.038 & Fe II 1121.97 & \nl
11 & 1096.94 & 0.202 & 0.047 & Fe II 1096.88 & \nl
12 & 1084.19 & 0.515 & 0.051 & N II 1083.99, N II* blend & \nl
13\tablenotemark{b} & 1066.77 & 0.135 & 0.047 & Ar I 1066.66 & \nl
14 & 1063.20 & 0.327 & 0.049 & Fe II 1063.18 + \htwo & \nl
15\tablenotemark{c} & 1048.19 & 0.096 & 0.035 & Ar I 1048.22 & 13.6 \nl
16 & 1039.18 & 0.433 & 0.044 & O I 1039.23 & \nl
17 & 1031.87 & 0.363 & 0.038 & O VI 1031.93 & \nl
18 & 1020.75 & 0.129 & 0.040 & Si II 1020.70 & \nl
19 & 1012.53 & 0.083 & 0.036 & S III 1012.50 & 14.4 \nl
20 & 1008.51 & 0.165 & 0.039 & NOID & 
\enddata

\tablenotetext{a}{Continuum determined with 2nd order local polynomial fit.}
\tablenotetext{b}{Up to 0.02 \AA\ may be contributed by \htwo\ ($J$~=~2).}
\tablenotetext{c}{Continuum may be depressed by adjacent weak lines.}

\end{deluxetable}

The identification of most of the strong interstellar features 
listed in Table 1 is secure \cite{M91}.
There are no obvious interstellar or intergalactic
candidates for the unidentified line at 1008.5 \AA, which may be spurious.
Although the observed features at 1066.8 and 1048.2 \AA\ 
presumably correspond to a pair of \arxi\ lines, their
equivalent widths at first glance appear to be anomalous 
(the 1048.2 \AA\ line should be stronger).  
As noted, however, the longer wavelength feature may be partially
blended with an \htwo\ line at 1066.9 \AA.
The shorter wavelength feature probably suffers from
local continuum depression by nearby \htwo.
Using the linear continuum shown in Figure 2, we
estimate a best fit equivalent width closer to 0.13 \AA.
These systematic effects, in combination with the 
statistical errors, do much to reconcile the anomaly
(especially if the features are on the flat part of
the curve-of-growth).

Below 1000 \AA, the increasing noise in the spectrum makes quantitative 
analysis of absorption features difficult,
and the continuum fitting routine fails.
Galactic features that are clearly detected include
a blend from \oxi\ (and \nxiii ?) near 989 \AA,
\cxiii\ (977.02 \AA), Lyman $\delta$ (972.54 \AA),
and an \nxi\ blend near 953.8 \AA.
Higher series Lyman lines are likely to be filled
in by diffuse emission (we correct for diffuse \hxi\ emission 
only through Lyman $\delta$).
An intriguing dip near 982 \AA\ may be associated
with \cxiii\ in the Virgo cluster, but this identification
is speculative.
Difficulties in continuum placement 
render the equivalent widths very uncertain; 
we attempt no quantitative analysis of absorption
lines below 1000 \AA.

\section{INTERSTELLAR GAS}

The equivalent widths of features at wavelengths of overlap
with the bandpass of the Hubble Space Telescope are consistent
with previous measurements from that instrument (\citeNP{MWSG91}, \citeNP{SLWMG93}).
For several other interstellar species (such as \fexii),
parameters inferred from the GHRS measurements \cite{SLWMG93}
yield predictions for lines in the ORFEUS band that are consistent
with the features observed.
Our spectrum represents the first detection
of \pxii, \fexiii, \arxi, and \sxiii\ along this sight line.
The statistical significance of these features is modest, as can 
be seen in Table 1.
The minimum columns inferred for these elements
(and for \nxi\ and \sixiii, features of which have been observed
previously but whose column densities have not been reported)
are shown in Table 1 and are of moderate interest only.  
Our minimum column for \fexiii\ is a factor of 2.5
below that established for \fexii\ \cite{SLWMG93}.
The ratio of \sxiii\ to \sxii\ \cite{SLWMG93} is at least 0.06 .

Interesting interstellar species unique to the ORFEUS
band include \htwo\ and \oxvi, on which we now focus our attention.
In Figure 2 we show a portion of the spectrum
between 1046 and 1066 \AA.  This region is comparatively
clear of competing interstellar lines and provides
a convenient hunting ground for features of \htwo.
The solid line is the observed spectrum; the lowest dotted
line is the 1 $\sigma$ error.
Overlying the data is a synthetic spectrum consisting 
of a flat continuum 
plus interstellar features convolved with the instrument response
(dashed line).  Wherever possible
we adopt column densities, velocities, and effective Doppler $b$-values
of atomic species
from \citeN{SLWMG93}.
An atomic-only model (omitted for clarity) does not reproduce the
observed modulation near 1050 \AA\ 
nor the observed width of the the strong \fexii\ line at 1063.2 \AA.

The total \htwo\ column density implied by our data
is about $1 \times 10^{15}$ cm$^{-2}$, 
corresponding to a molecular fraction 
($f_{H_2}$) of about $2 \times 10^{-5}$.  This value is 
not exceptional for a low-reddening sight line \cite{SB82}.
The rotational excitation temperature $T_{01}$ is 
poorly constrained but appears to be consistent with
the range typical of interstellar clouds (50 -- 80 K).

In Figure 3 we show a portion of the spectrum
between 1023 and 1041 \AA.
The legend is as in Figure 2, except that we have
not labeled a few weak features to avoid crowding.
Features labeled Cld. 1 and Cld. 2 arise in intergalactic gas.
The discrepancy between their synthetic and observed spectra
is discussed below.
The strong galactic Lyman $\beta$ line is well fit 
in the wings; at the core, the subtraction of bright diffuse
emission introduces a significant uncertainty 
indicated by the large peak in the error tracing.
The stronger \oxvi\ doublet member (1031.9 \AA) is resolved 
cleanly; the weaker member is blended but clearly present.
To produce the fit shown in Figure 3,
we found it necessary to set the column density of \cxii*
near the upper limit from \citeN{SLWMG93}.  This modification
to the \cxii* column does not significantly affect the 
previous determination that the cooling rate per
nucleon is substantially below (about 1/6 of)
the Galactic average \cite{SLWMG93}.

With a $b$ value of about 35 \kms\ (estimated from the \nxv\
profile of \citeNP{SST97}), our best estimate
of the \oxvi\ column density is $7 \pm 2 \times 10^{14}$ cm$^{-2}$.
\citeN{SST97} found an \nxv\ column of
$8 \times 10^{13}$ cm$^{-2}$, yielding an \oxvi / \nxv\ ratio of about 10.
The ratio along disk sight lines 
varies from about 10 to 20 (compilation of \citeNP{HB96}).
Ionization in this component of the hot galactic gas toward 3C273 
therefore appears to be roughly similar to corresponding conditions
in the disk.  The fact that our measurement is near the
low end of the disk range is qualitatively consistent with 
a scenario in which the gas is heated impulsively 
then cools as it moves away from the plane.
Whether the comparatively high total column density
is significantly nonrepresentative of the high latitude sky
\cite{HB96} will be determined
with certainty only after study of many additional sight lines.
The presence of Radio Loops I and IV \cite{BHS71} 
cautions at the
very least that division of our projected column by
the midplane density \cite{J178} should not be taken 
as a reliable estimate of the typical scale height.
However, the fact that this first clear measurement of 
columns for both \oxvi\ and \nxv\ yields a value
comparable to that in the disk bodes well 
for future far-ultraviolet missions such as FUSE.
The previous measurement of \oxvi\ toward 3C273 \cite{D93}
had set only a lower limit on the column density, corresponding
to an \oxvi / \nxv\ ratio substantially below the range of disk values.

\section{INTERGALACTIC GAS}

Sixteen otherwise unidentified absorption features in the GHRS spectrum
were attributed to intergalactic Lyman $\alpha$ clouds by \citeN{MWSG91}.
Most of these are sufficiently weak that their higher
Lyman series lines would not be detectable in our spectrum.
We do find probable Lyman $\beta$ counterparts to the
strongest proposed Lyman $\alpha$ forest features, however.
\citeN{MWSG91} propose a cloud at $cz$ = 19900 \kms;
from its column density and $b$ value we predict a 
Lyman $\beta$ equivalent width of 0.080 \AA\ at 1093.93 \AA.
We find an otherwise unidentified
feature within 0.2 \AA\ of the expected position 
with an equivalent width of 0.060 $\pm$ 0.038 \AA\
confirming, at least to 1.6 $\sigma$, the intergalactic \hxi\ origin.

Of greater interest are a strong pair of features attributed to
comparatively low redshift \hxi\ gas in the Virgo supercluster.
Their Lyman $\alpha$ lines appear near +1000 and +1600 \kms\ \cite{WRWMH95}.
Lyman $\beta$ counterparts are expected at 1029.21 and 1031.16 \AA.
We detect absorption features at 1029.11 and 1031.14 \AA;
however, the equivalent widths are stronger than expected.
\citeN{WRWMH95} employ profile fitting to
constrain column densities and Doppler $b$ values.
Based on those parameters, the 
Lyman $\beta$ equivalent widths are predicted to be 0.098 
and 0.10 \AA, respectively.
We observe 0.145 ($\pm$ 0.037) and 0.241 ($\pm$ 0.032) \AA.
The disagreement is seen clearly in Figure 3, where the
synthetic spectrum relies on the \citeN{WRWMH95} parameters.

The discrepancy is particularly pronounced for the +1600 \kms\ cloud.
Even with the parameters proposed by \citeANP{WRWMH95} the
Lyman $\alpha$ line is saturated 
($\tau_0$~=~3.7), so determination of a column
density based solely on profile fitting of that feature may be
affected by unresolved velocity structure and/or uncertainties 
in the instrumental profile.
We have analyzed the GHRS spectrum to extract
the equivalent width of the Lyman $\alpha$ line.
In combination with the Lyman $\beta$ line from the ORFEUS
spectrum, we establish constraints
on the logarithmic column density and effective $b$ value
for the +1600 \kms\ cloud using a curve-of-growth technique.
We show the results, and the parameters proposed
by \citeANP{WRWMH95}, in Figure 4.  
Our results suggest that if the feature arises
in a single cloud, its column density is higher
by at least a factor of four compared to the 
value suggested by \citeANP{WRWMH95}.
The curve-of-growth results are consistent with $b$ values as low 
as 12 \kms\ and column densities in excess of $10^{18}$ cm$^{-2}$.
Although this extremum of parameter space is probably
inconsistent with the Lyman $\alpha$ profile fitting,
it is not difficult to construct a multiple-cloud
system that closely approximates the theoretical
Voigt profile inferred from the \citeANP{WRWMH95} parameters
while hiding a great deal of gas in the saturated core.

\citeN{GBJS93} searched for 21 cm emission from
Virgo gas near the velocities of the absorbing clouds, 
detecting no \hxi\ to a limit of about $2.8 \times 10^{19}$ cm$^{-2}$.
By adopting a relationship between the 
volume density of clouds with columns near
the radio limit and of clouds with columns near
those of the absorption line systems \cite{T87},
the radio survey constrains the minimum size of the 
absorbing systems.
The minimum permitted cloud size grows slowly with
column density ($r_{cloud} \propto N^{0.25}$),
so our results increase the minimum cloud size
from about 4 kpc \cite{GBJS93}
only to about 6 kpc, at least for the smallest 
columns within our 95\% confidence interval.

Other interstellar species that might conceivably be 
blended with the +1600 \kms\ cloud include \clxi\ at 1030.88 \AA,
or \htwo\ ($J$~=~3) at 1030.89 \AA.  Based on the
nondetection of other features from these species,
we conservatively limit the equivalent width of
their combined contribution to no more than 0.050 \AA.
A contribution from Galactic \oxvi\ at high negative velocity is 
more difficult to rule out formally from our spectrum.
Neither \nxv\ nor \cxiv\ appears at high negative velocity, however \cite{SST97}.

\section{CONCLUSIONS}
\label{conclusions}

We observed the bright quasi-stellar object 3C273
for 10,797 s with the Berkeley spectrograph aboard 
the \orf-II mission \cite{Hetal98}.
The resulting spectrum offers intermediate spectral resolution
($\lambda$ / 3000 FWHM) and a 1$\sigma$ noise corresponding
to an absorption feature with equivalent width of about 0.040 \AA.

The spectrum reveals, at modest
statistical significance, a variety of previously undetected
low ionization interstellar species, including
\pxii, \fexiii, \arxi, and \sxiii.
There is marginally significant evidence for \htwo,
with a total column density of about $10^{15}$ cm$^{-2}$.
The $J$~=~0 and $J$~=~1 rotational levels seem
comparably populated, consistent with an
excitation temperature $T_{01}$ 
typical for interstellar clouds (50 -- 80 K).

We detect the stronger interstellar \oxvi\ doublet
member unambiguously; the weaker member is blended.
If the Doppler $b$ value for \oxvi\ is comparable
to that determined for \nxv\ \cite{SST97},
the \oxvi\ column density is $7 \pm 2 \times 10^{14}$ cm$^{-2}$.
The \oxvi / \nxv\ ratio is about 10, consistent with
the low end of the range observed in the 
disk (compilation of \citeNP{HB96}).
If impulsively heated gas cools as it
moves away from the Galactic disk, the \oxvi / \nxv\ ratio 
might be expected to decrease (compared to the disk value)
along halo sight lines.

We detect Lyman $\beta$ counterparts to intergalactic Lyman $\alpha$
features identified by \citeN{MWSG91} at $cz$ = 19900, 1600, and 1000 \kms.
Counterparts to other putative Lyman $\alpha$ clouds
along the sight line are below our detection limit.
The strengths of the two very low redshift Lyman $\beta$ features,
which are believed to arise in Virgo intracluster gas,
exceed preflight expectations \cite{WRWMH95}.
A curve-of-growth analysis sets a minimum \hxi\ column density
of $4 \times 10^{14}$ cm$^{-2}$ for the 1600 \kms\ cloud.
Following \citeN{GBJS93}, our revised column density
and nondetection of 21 cm radio emission discussed in
that work can set a lower limit of about 6 kpc to the 
size of the Virgo cluster clouds.

\acknowledgments

We acknowledge our colleagues on the
\orf\ team and the many NASA and DARA personnel who helped make the
\orf-II mission successful. This work is supported by NASA grant
NAG5-696.

\clearpage 





\newpage 

\begin{figure}
\epsscale{0.8}
\plotone{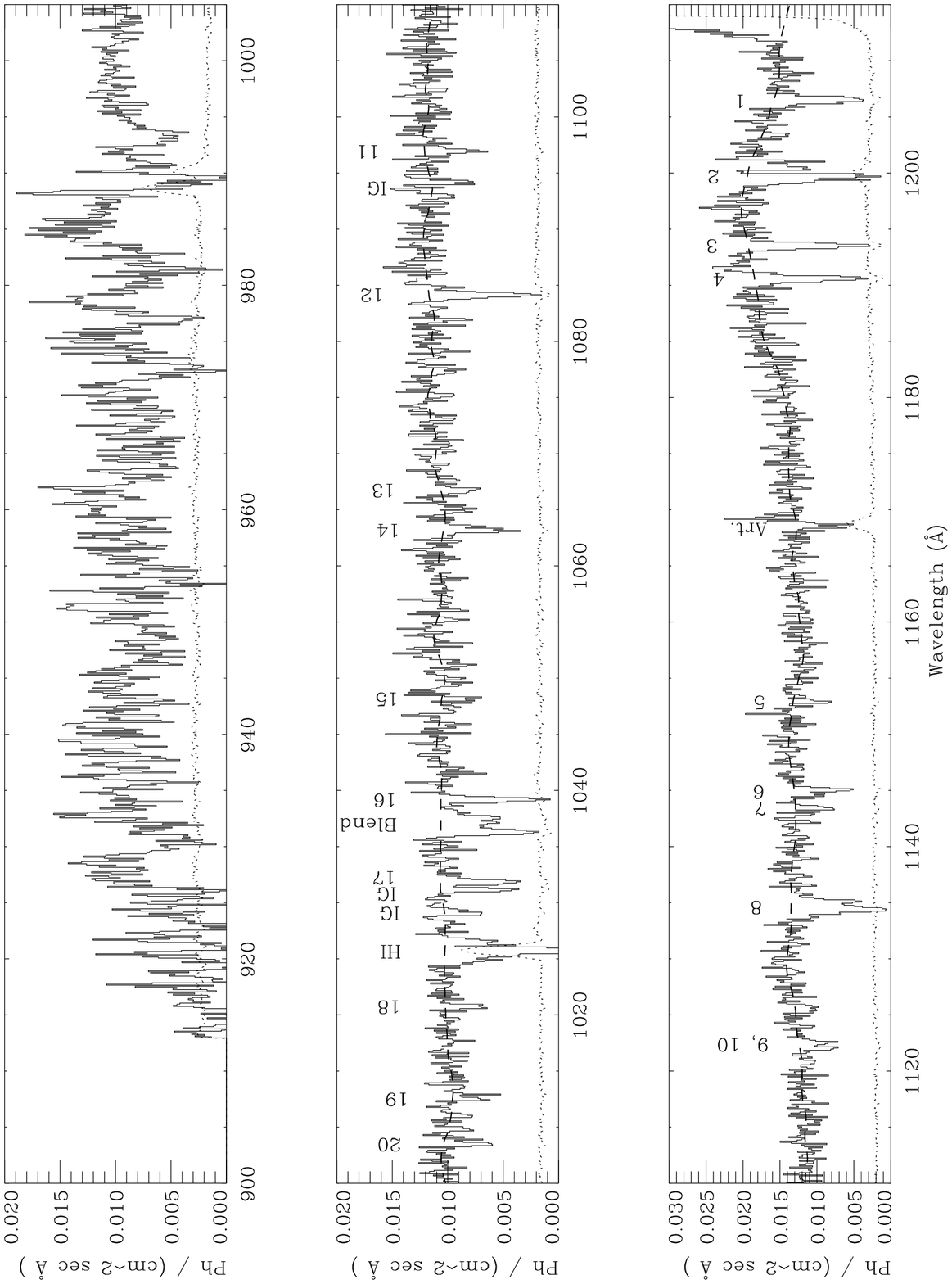}
\caption{Observed spectrum of 3C 273 (solid line), 1 $\sigma$ 
uncertainty (dotted line), continuum fit longward 
of 1000 \AA (dashed line).
Numbered interstellar absorption features are
identified in Table 1.  The modulation near 1069 \AA\ is
an artifact associated with second-order \hexi\ diffuse emission.
``IG'' indicates intergalactic Lyman $\beta$ features.
\label{bigfig}
}
\end{figure}

\newpage 

\begin{figure}
\plotone{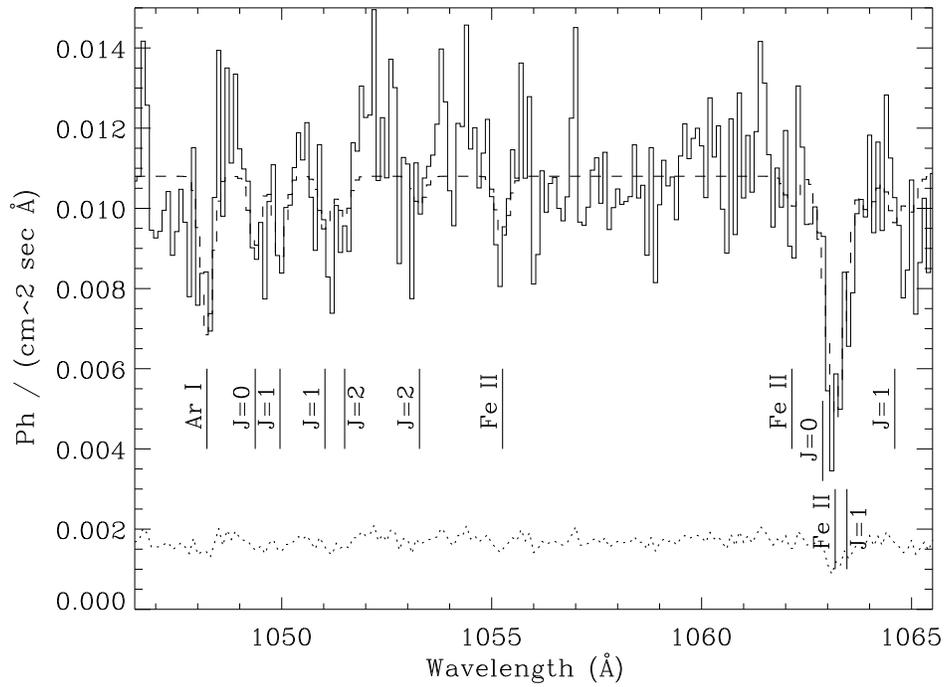}
\caption{Observed spectrum of 3C 273 (solid line) with 1 $\sigma$ 
uncertainty (lower dotted line).  Synthetic spectra overlying
the data are based on a flat continuum plus
interstellar atomic and \htwo\ features (dashed line).
\htwo\ features are labeled with $J$ level of lower state.
\label{h2fig}
}
\end{figure}

\newpage 

\begin{figure}
\plotone{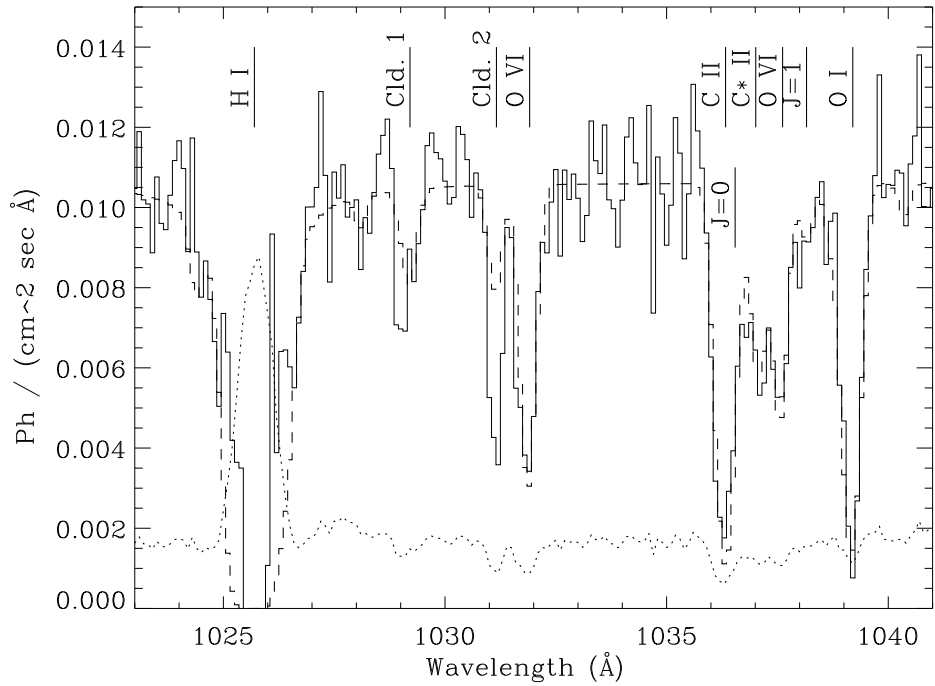}
\caption{Observed spectrum of 3C 273 (solid line) with 1 $\sigma$ 
uncertainty (dotted line).  Synthetic spectrum overlying
the data is based on a flat continuum plus
interstellar atomic and \htwo\ features and intergalactic features.
\htwo\ features are labeled with $J$ level of lower state.
Features labeled ``Cld. 1'' and ``Cld. 2'' are Lyman $\beta$
clouds in the Virgo cluster.
\label{ovifig}
}
\end{figure}

\newpage 

\begin{figure}
\plotone{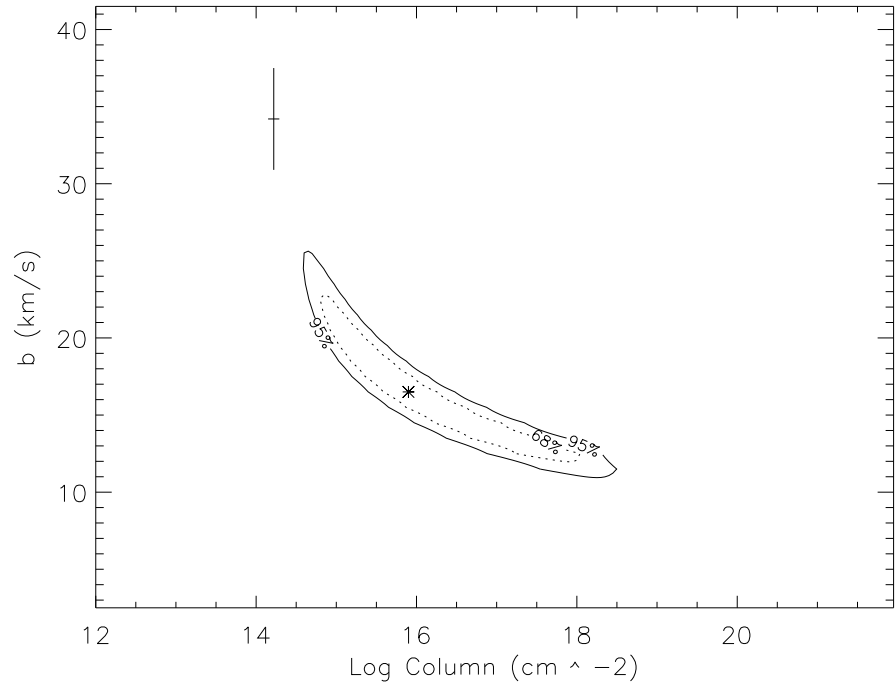}
\caption{Constraints on logarithmic column
density and effective Doppler $b$ value
for the 1600 \kms\ \hxi\ cloud.
The cross centered at (14.22, 34) is based on
Lyman $\alpha$ profile fitting (Weymann et al. 1995).
The best-fit, 68\% and 95\% confidence
intervals based on the equivalent widths of 
Lyman $\alpha$ (GHRS) and $\beta$ (ORFEUS)
are also shown.
\label{nbfig}
}
\end{figure}

\end{document}